\documentclass[pre,twocolumn,showpacs]{revtex4}
\usepackage{graphicx}
\usepackage{amsmath,amssymb}
\usepackage[usenames, dvipsnames]{color}
\begin{document}
\title{Ionic and electronic transport properties in dense plasmas\\by orbital-free density functional theory}
\author{Travis Sjostrom and J\'er\^ome Daligault}
\affiliation{Theoretical Division, Los Alamos National Laboratory,
Los Alamos, New Mexico 87545}
\date{\today}
\begin{abstract}
We validate the application of our recent orbital-free density functional theory (DFT) approach, [Phys. Rev. Lett. 113, 155006 (2014)], for the calculation of ionic and electronic transport properties of dense plasmas.
To this end, we calculate the self-diffusion coefficient, the viscosity coefficient,  the electrical and thermal conductivities, and the reflectivity coefficient of hydrogen and aluminum plasmas.
Very good agreement is found with orbital-based Kohn-Sham DFT calculations at lower temperatures.
Because the method does not scale with temperature, we can produce results at much higher temperatures than is accessible by the Kohn-Sham method. 
Our results for warm dense aluminum at solid density are inconsistent with the recent experimental results reported by Sperling {\it et al.} [Phys. Rev. Lett. 115, 115001 (2015)].
\end{abstract}
\maketitle

\section{Introduction}

Recently we presented an approach for ``Fast and accurate quantum molecular dynamics of dense plasmas across temperature regimes'' based on a carefully designed orbital-free implementation of density functional theory (DFT) \cite{sjostromdaligault}.
Our orbital-free approximation retains the accuracy of the orbital-based Kohn-Sham method (it reproduces the electron density to high accuracy), while being computationally less expensive and reaching much higher temperatures than are accessible with the Kohn-Sham method.
This was shown in \cite{sjostromdaligault} for {\em static} properties, including the equation of state of hydrogen from 1 to 100 eV, as well as for the pair distribution functions of aluminum near melt and in warm dense matter conditions.

In this paper, we extend our study to the calculation of {\em dynamical} properties.
To this end we calculate both ionic and electronic transport coefficients, including the ion self-diffusion coefficient, the ion shear viscosity coefficient, the electrical and thermal conductivities and the reflectivity coefficient.
As with the static properties, we find very good agreement with Kohn-Sham DFT calculations. 
Moreover, the orbital-free approach provides significant relief from the computational cost temperature bottleneck of the Kohn-Sham method, allowing us to calculate accurately more extreme conditions.
This is particularly useful for the calculation of ionic transport properties, which necessitate much longer simulations than necessary for equation of state calculations.

The paper is organized as follows.
In section \ref{ionictransport}, we focus on ionic transport properties and validate our method for the calculation of the self-diffusion and viscosity coefficients in hydrogen and aluminum plasmas by comparison with Kohn-Sham method at low temperature, and then extend those calculations to higher temperatures. 
In section \ref{electronictransport},  we focus on electronic transport properties and consider the  electrical and thermal conductivities of hydrogen and aluminium plasmas; here, we make use of both the ionic positions and associated electron density from the orbital-free calculations as a shortcut to the orbitals required in the Kubo-Greenwood formalism.
In both sections, we systematically compare the computational cost of our orbital-free approach with the Kohn-Sham method.
Finally, special attention is given to the electrical conductivity of warm dense aluminum at 2.7 $\rm g/cm^3$ in view of the recent experimental determination at the Linac Coherent Light Source facility reported in \cite{Sperling}.
We find the latter experimental conductivity results to be inconsistent with our orbital-free DFT results, for which we provide further discussion.

\section{Ionic transport} \label{ionictransport}

The full details of the orbital-free formulation and implementation may be found in Ref. \cite{sjostromdaligault} and its Supplemental Material.
Applying that approach, here we calculate the ion self-diffusion and shear viscosity coefficients of hydrogen and aluminum plasmas using the standard Green-Kubo relations \cite{Hansen}. 
Hence the self-diffusion coefficient is obtained from the time integration of the velocity autocorrelation function,
\begin{align}
D = \frac{1}{3N}\sum_{i=1}^N{ \int_0^{\infty} \left< \mathbf{V}_i(t) \cdot \mathbf{V}_i(0) \right> dt }\;, \label{selfdiffusion}
\end{align}
where $N$ is the total number of ions in the simulation cell, $\mathbf{V}_i(t)$ is the velocity of the $i^{th}$ ion at time $t$, and the brackets indicate the equilibrium thermal average.
The ionic shear viscosity coefficient is obtained by integrating the autocorrelation function of the off-diagonal component of the stress tensor,
\begin{align}
  \eta = \frac{1}{6\Omega k_B T}\sum_{\alpha\neq\beta}^3{ \int_0^{\infty} \left< \sigma_{\alpha,\beta}(t)\sigma_{\alpha,\beta}(0) \right> dt }\;,
\end{align}
where $\Omega$ and $T$ are the system volume and temperature, $k_B$ the Boltzmann constant, and $\sigma$ is the ionic stress tensor, given by 
\begin{align}
  \sigma = \sum_{i=1}^N m \mathbf{V}_{i} \mathbf{V}_{i} + \sigma^{II} + \sigma^{IE}\;.
\end{align}
The first term on the right hand side is the kinetic contribution, while the ion-ion, $\sigma^{II}$,and ion-electron, $\sigma^{IE}$, potential contributions are as given in Ref. \cite{profess}, and depend only on the electron density and ion positions.

\begin{figure}
  \centering
  \includegraphics[width=\columnwidth]{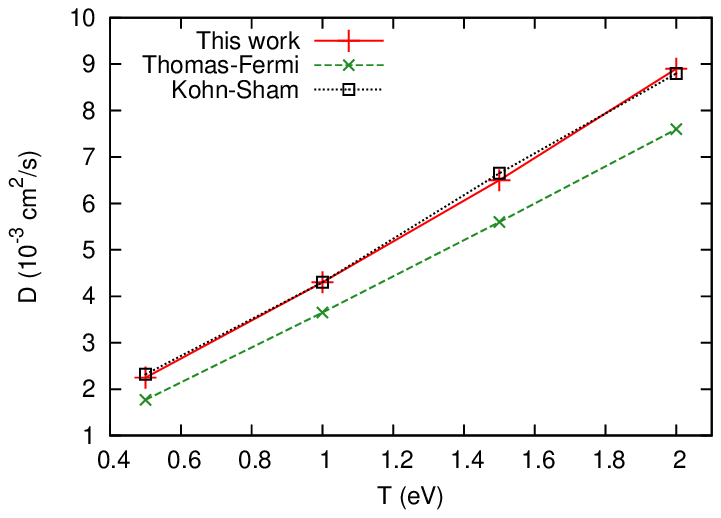}
  \includegraphics[width=\columnwidth]{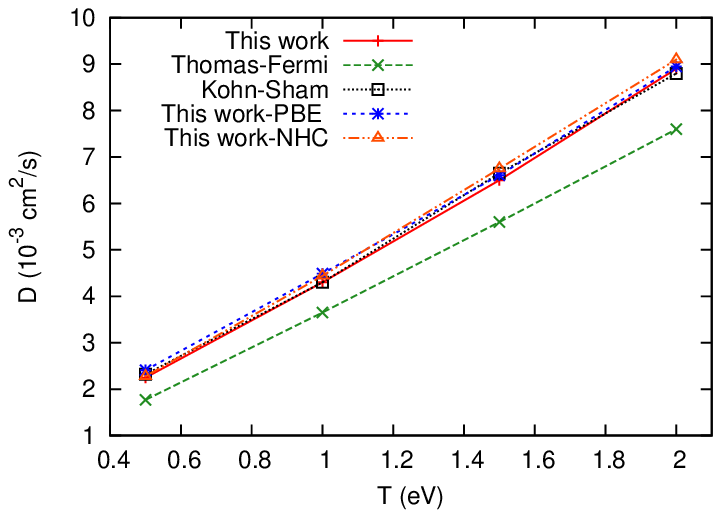}
  \caption{Self-diffusion coefficient of hydrogen plasmas at 2 g/cm$^3$ and lower temperatures, where comparison with Kohn-Sham is possible. Above, our results show excellent agreement with Kohn-Sham while Thomas-Fermi approach exhibits $\sim$20-30\% difference. Below, we see that for hydrogen at these temperatures and densities, there is minimal difference in using LDA or PBE exchange-correlation functionals. Also using a Nose-Hoover chain thermostat \cite{nhc}, we find negligible difference with the isokinetic thermostat used in all other calculations discussed in this paper.}
  \label{fig:Htest}
\end{figure}

\subsection{Hydrogen}

\begin{figure}
  \centering
  \includegraphics[width=\columnwidth]{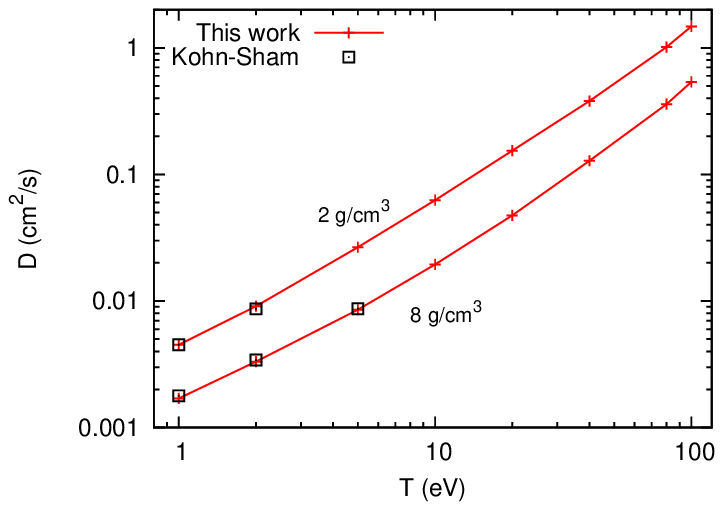}
  \includegraphics[width=\columnwidth]{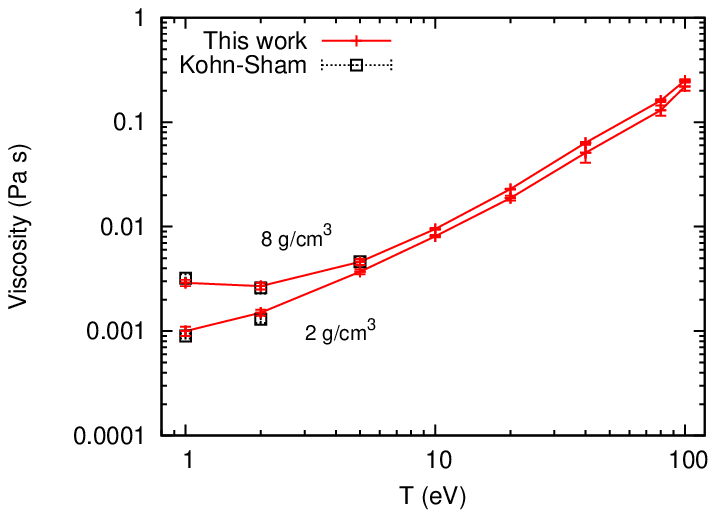}
  \caption{Self-diffusion coefficient (upper panel)  and shear viscosity coefficient  (lower panel) for hydrogen plasmas at 2 and 8 g/cm$^3$in the range of temperatures from 1 eV to 100 eV. Very good agreement is shown with Kohn-Sham at lower temperatures.}
  \label{fig:Hdiff}
\end{figure} 
The Kohn-Sham calculations were performed using the Quantum-\textsc{Espresso} code\cite{QE} at the $\Gamma$-point only and used a projector augmented-wave (PAW) pseudopotential, while the orbital-free calculations employed a local pseudopotential as described in Ref. \cite{sjostromdaligault}. 
For comparison, we also show results obtained using the cruder Thomas-Fermi approximation.
In all cases, the simulations included 128 atoms in the unit cell, with time steps from 0.002-0.2 fs depending on temperature, and they were performed in the isokinetic ensemble\cite{Bussietal}.
In most calculations, the local density approximation (LDA) \cite{PerdewZunger} for the exchange-correlation energy was used; other calculations were done using the Perdew-Burke-Ernzerhof (PBE) generalized gradient approximation \cite{pbe}.
The orbital-free calculations were equilibrated for 10,000 steps and the statistics gather for 120,000 steps, while for the Kohn-Sham calculation 40,000 steps were completed after equilibration.

Figure \ref{fig:Htest} shows results at density 2 g/cm$^3$ and lower temperatures for which comparative Kohn-Sham results can be obtained.
In the upper panel our results are shown to agree very well with the Kohn-Sham results \cite{noteonwang}.
As for the Thomas-Fermi approximation, it underestimates the self-diffusion by 20-30\% at these conditions.

We note that the calculations were run on the same 48-core single node hardware, and the time to complete 40,000 molecular dynamics steps at $T=2$ eV for the Kohn-Sham and our orbital-free methods were 87.25 and 7.35 hours respectively, giving a nearly twelve times speed up for the orbital-free case. 
For still increasing temperature, the Kohn-Sham approach costs more in machine memory and cpu time, while the orbital-free method has no increased cost with increasing temperature.

This lack of temperature scaling allows us to extend our hydrogen results to much higher temperatures.
In Fig. \ref{fig:Hdiff}, we show the self-diffusion (upper panel) and shear viscosity (lower panel) coefficients for hydrogen at 2 and 8 g/cm$^3$ in the range of temperatures from 1 eV to 100 eV.
At the lowest temperatures of 1 and 2 eV ( and 5 eV for 8 g/cm$^3$) Kohn-Sham calculations are also shown for comparison and we see an agreement to within 4\% of our method.

The self-diffusion results in general have an uncertainty of 5\%, which is determined by inspection of the convergence of the Kubo relation (\ref{selfdiffusion}).
The Thomas-Fermi approach, however, differs from our results by 15-20\% up to 10 eV, and then gradually comes in to agreement with our results, showing a difference within 1\% at 80 eV.
For the viscosity coefficient, the orbital-free results agree with the Kohn-Sham results to a maximum error of 10\%, which is within the uncertainties of the calculations which are 10-12\%.
While the viscosity coefficients at 2 g/cm$^3$ increase monotonically in this temperature region, the 8 g/cm$^3$ case exhibits a minimum around 2 eV.
This is indicative of the transition from a weakly coupled to a strongly coupled plasma regime as the temperature decreases, similar to that found in the simpler one-component plasma model \cite{ocp}.

\subsection{Aluminum}

\begin{figure}[t]
  \centering
  \includegraphics[width=\columnwidth]{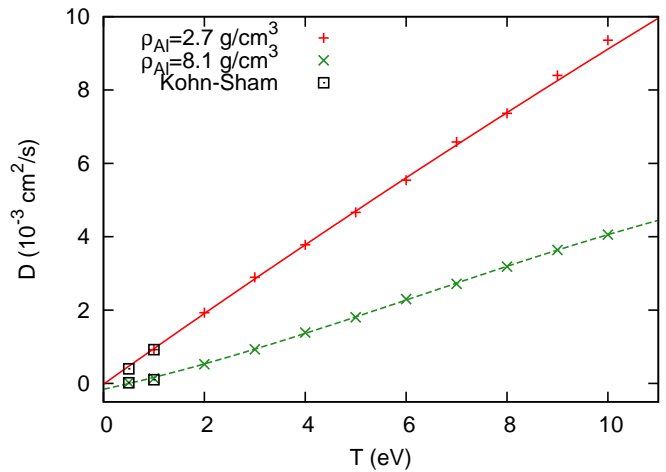}
  \caption{Self-diffusion coefficient for aluminum plasmas at 2.7 and 8.1 g/cm$^3$ and from 0.5 to 10 eV. Again good agreement is shown with Kohn-Sham calculations at 0.5 and 1 eV. A smooth fit to the orbital-free calculations is shown.}
  \label{fig:Aldiff}
\end{figure} 
The Kohn-Sham calculations were performed at the $\Gamma$-point using 64 atoms in the unit cell, and used a 680 eV plane wave cutoff.
A total of 180 bands were calculated to achieve a 10$^{-3}$ threshold in the occupation number at 2.7 g/cm$^3$ and 1 eV, and 160 bands were needed for a maximum occupation of $2\times10^{-4}$ at 8.1 g/cm$^3$ and 1 eV.
The orbital-free calculations were performed for 108 atoms on a 64$^3$ grid. 
Finally, the local pseudopotential for the orbital-free calculations \cite{sjostromdaligault} and the PAW for the Kohn-Sham calculations include 3 valence electrons only; this limits the the maximum temperature permissible to about 10 eV.

Figure \ref{fig:Aldiff} shows the self-diffusion results for warm dense aluminum at ambient solid density, 2.7 g/cm$^3$, and three times compression, 8.1 g/cm$^3$, in the range of temperature from 0.5 to 10 eV.
As with hydrogen we see very good agreement in the self-diffusion coefficient at lower temperature where the calculations overlap. 
However, the time for the completion of 45,000 molecular dynamics timesteps for the Kohn-Sham calculation at 2.7 g/cm$^3$ and 1 eV was 225 hours on a 48 core node, while the time to complete the same number of steps on just 16 cores of the same machine in the orbital-free case was 13.1 hours.

\begin{figure}[t]
  \centering
    \includegraphics[width=\columnwidth]{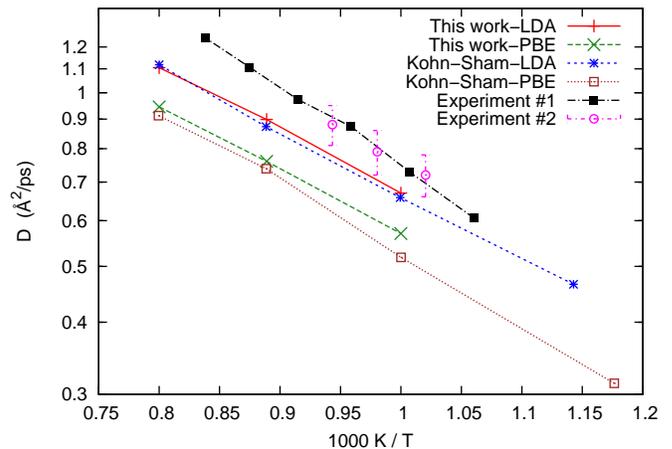}
  \caption{Self-diffusion coefficient for liquid aluminum near melt compared with experimental results. Our results are close to those for Kohn-Sham for both PBE and LDA. None of the simulations are in agreement with Experiment \#1 \cite{exp1} and Experiment \#2 \cite{exp2} suggesting inaccuracy due to the exchange-correlation functional (for more discussion, see \cite{JaksePasturel})}
  \label{fig:liqAl}
\end{figure}
Finally we consider the case of liquid aluminum near melt.
In our previous work \cite{sjostromdaligault} we showed excellent agreement for the ion-ion pair distribution function as compared with both Kohn-Sham method and experimental results.
Following that, we examine here the self-diffusion as calculated at the experimental densities and temperatures \cite{Assaeletal}.
In Fig. \ref{fig:liqAl} our orbital-free results are plotted using both LDA and PBE exchange-correlation functionals, and are compared with the the previous Kohn-Sham results of \cite{JaksePasturel,notepasturel}.
In each case, the orbital-free and Kohn-Sham results agree closely, although the LDA agreement is somewhat better than the PBE agreement.
We believe that this is due to lower accuracy in the PBE local pseudopotential.

Here, as opposed to the hydrogen case of Fig.~\ref{fig:Htest}, there is significant difference between the exchange-correlation functionals, and further neither LDA nor PBE agree with the recent experimental self-diffusion results \cite{exp1,exp2} also shown in Fig. \ref{fig:liqAl} (for more discussion, see \cite{JaksePasturel}).
It is of note our orbital-free approach captures these differences between exchange-correlation functionals, which does highlight the accuracy of our kinetic functional, as the kinetic energy contribution is typically an order of magnitude larger than the exchange-correlation contribution \cite{notegonzales}.

\section{Electronic transport} \label{electronictransport}

So far we have been able to express the ionic transport properties completely within the framework of orbital-free DFT, which involves the electron density only.
However when considering electronic transport coefficients, transitions between quantum states of the system must be considered and it is necessary to return to a description in terms of Kohn-Sham orbitals.
To this end, our approach is to first perform an orbital-free calculation from which ionic configurations and corresponding electron densities are selected at a subset of time steps along the simulation.
(Recall that in \cite{sjostromdaligault} we showed that the densities obtained using our orbital-free approach are in very good agreement with those obtained with a self-consistent Kohn-Sham calculation.)
Then, for each ionic configuration and electron density, the Kohn-Sham potential is readily calculated and the single-particle Kohn-Sham spectrum of eigenstates is calculated by diagonalization of the single-particle Kohn-Sham equation.

The spectrum is then used to calculate the electrical $\sigma$ and thermal $\kappa$ conductivities by evaluating the Kubo-Greenwood formula (see \cite{Holstetal} and references therein),
\begin{align}
  \sigma=L_{11}\;, \quad \kappa=\frac{1}{T} \left( \L_{22}-\frac{L_{12}^2}{L_{11}}\right)\;,
\label{eq:etrans}
\end{align}
where the frequency-dependent Onsager coefficients are given by
\begin{align}
  L_{mn}(\omega) = &\frac{2\pi e^{4-m-n}}{3 V m_e^2 \omega}
  \sum_{\mathbf{k}\nu\mu} \left | \left< \mathbf{k}\nu | \mathbf{\hat{p}} | \mathbf{k}\mu \right> \right|^2 (f_{\mathbf{k}\nu} - f_{\mathbf{k}\mu}) \nonumber \\ 
& \left( \frac{\varepsilon_{\mathbf{k}\nu}+\varepsilon_{\mathbf{k}\mu} }{2} - h_e\right)^{m+n-2}
\delta(\varepsilon_{\mathbf{k}\mu}-\varepsilon_{\mathbf{k}\nu}-\hbar\omega)\;.
\label{eq:Ons}
\end{align}
Here $\mathbf{\hat{p}}$ is the momentum operator, $\mathbf{k}$ is the specific $k$-point in the Brillouin zone, $\mu$ and $\nu$ label the band index, $\varepsilon$ are the single-particle energies, and $f$ are the single-particle occupations determined through the Fermi-Dirac distribution. Additionally $e$ and $m_e$ are the electron charge and mass and $h_e$ is the average electron enthalpy per electron. Lastly is the $\delta$ function, which is approximated by a Lorentzian function (see details in Appendix \ref{appendix1}) due to the discreteness of the energy levels. 

\begin{figure}[t]
  \centering
  \includegraphics[width=\columnwidth]{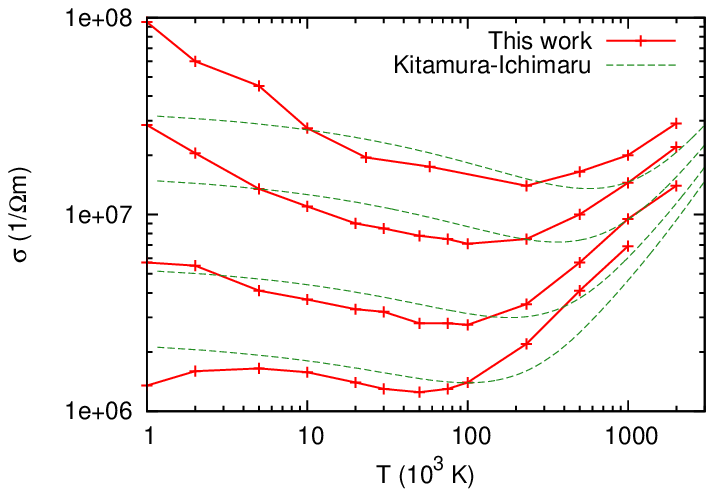}
  \includegraphics[width=\columnwidth]{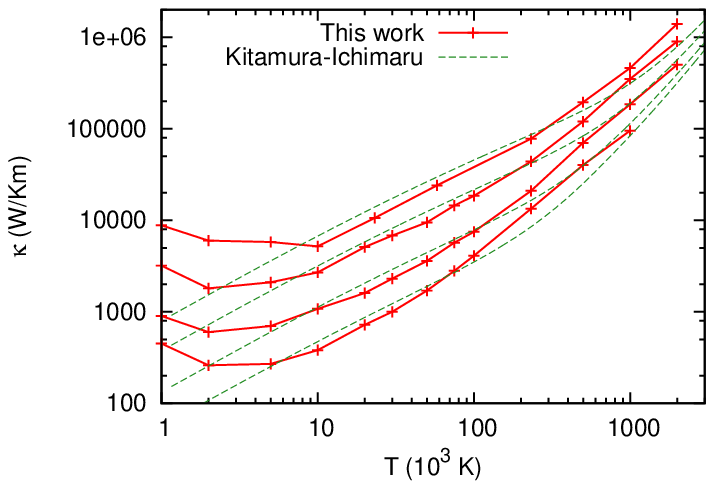}
  \caption{Electrical (upper panel) and thermal (lower panel) conductivities for hydrogen at 10, 5, 2, 1 g/cm$^3$  (decreasing from top to bottom curves) from 1,000- 2,000,000 K. For reference, we compare with the prediction of the Kitamura-Ichimaru model \cite{KitamuraIchimaru1995}.}
\label{fig:Hcond}
\end{figure}
Finally, for each ionic configuration, $\sigma$ and $\kappa$ are calculated and then averaged over all sampled configurations to obtain a converged result.
The reflectivity coefficient $R$ is then obtained from the frequency-dependent electrical conductivity $\sigma(\omega)$ (see appendix \ref{appendix2}).

The main difference of our approach from a full Kohn-Sham calculation is that we use the density from the orbital-free calculation in order to calculate the Kohn-Sham potential and then find the single-particle orbitals and occupation numbers by a single diagonalization as opposed to a fully self-consistent Kohn-Sham calculation which may require 10-20 diagonalization steps or more depending on the initial guess. This then decreases our computation time by at least an order of magnitude \cite{notelambertetal}. 

\subsection{Hydrogen}

\begin{figure}[t]
  \centering
  \includegraphics[scale=0.9]{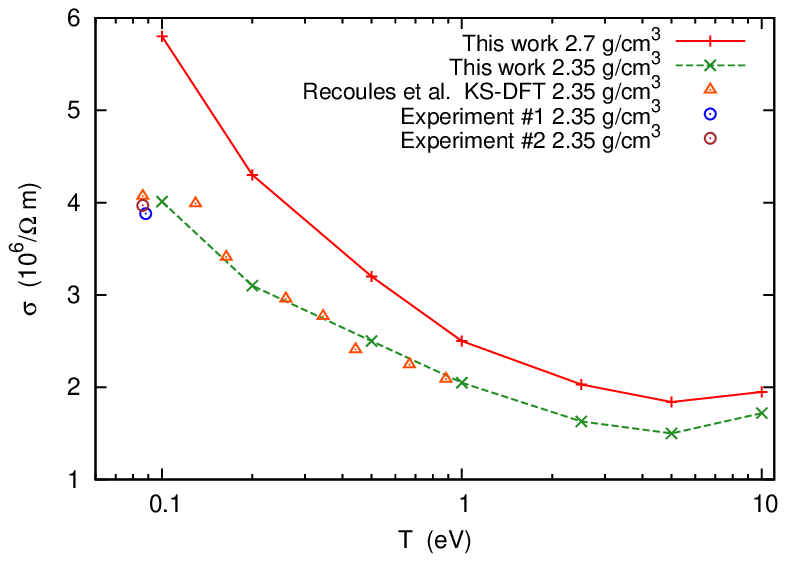}\\
  \includegraphics[scale=0.9]{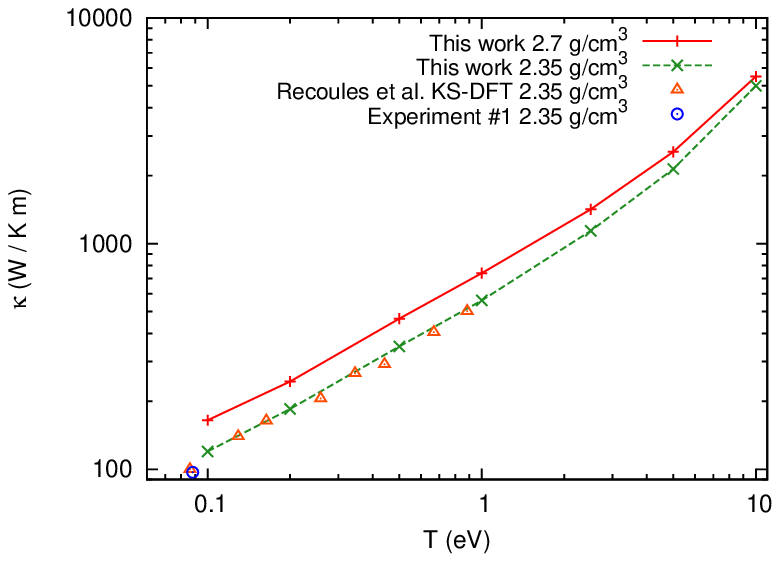}\\
  \includegraphics[scale=0.9]{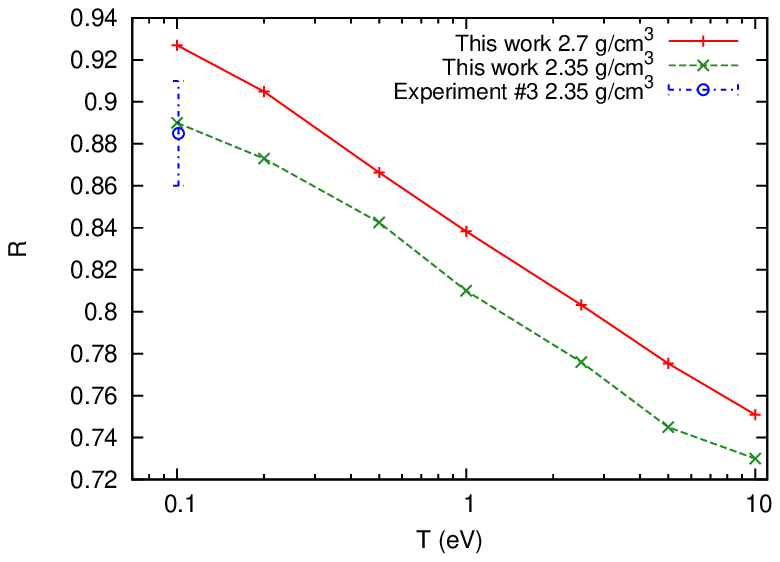}
  \caption{Electrical conductivity, thermal conductivity and reflectivity coefficient of aluminum plasmas at 2.7 and 2.35 g/cm$^3$. Previous Kohn-Sham and experimental results at the liquid density agree with our calculations.}
\label{fig:Alcond}
\end{figure} 
We have performed calculations for hydrogen plasmas from 1-10 g/cm$^3$ and temperatures from 1,000-2,000,000 K.
As discussed in \cite{Lambertetal} convergence with respect to the number ions, the number of $k$-points and the width of the Lorentzian smearing is critical.
We have used 96 atoms for all calculations, except at 2,000,000 K where we used 40 atoms. At 1,000,000 K we performed calculations with 40 and 96 atoms and found no difference in results.
For higher densities and lower temperatures a $3^3$ $k$-grid was required, moving to higher temperatures a $2^3$ grid and finally $\Gamma$-point only calculations were sufficient.
A Lorentzian smearing of 0.1 eV was also adequate (see Appendix \ref{appendix1}).

The results for the electrical and thermal conductivities $\sigma$ and $\kappa$ are plotted in Fig. \ref{fig:Hcond}.
We find that our results agree well with the previous calculations of Ref. \cite{Holstetal} and Ref. \cite{Lambertetal} for both $\sigma$ and $\kappa$.
We also find good agreement with the deuterium results (at twice mass density) of Ref. \cite{Huetal} for $\kappa$.
Here though we are able to complete calculations in the direction of lower densities and higher temperatures than in those previous works which is where the Kohn-Sham method becomes more computationally expensive.

\subsection{Aluminum}

We have calculated the electrical and thermal conductivities and the reflectivity coefficient of warm dense aluminum from 0.1-10 eV at solid density (2.7 g/cm$^3$) and at the ambient liquid melt density of 2.35 g/cm$^3$.
Here we found a much smaller width in the Lorentzian smearing, 0.015 eV, was required for convergence of the calculations which all used 64 atoms and a $2^3$ $k$-grid except at 10 eV which used only the $\Gamma$-point.
Also the 3-electron local pseudopotential used here limits us to temperatures less than 10 eV.

The results of our calculations are shown in Fig. \ref{fig:Alcond} together with previous results, including three independent experimental measurements.

Our orbital-free results at 2.35 g/cm$^3$ for $\sigma$ and $\kappa$ are in very good agreement with previous Kohn-Sham calculations of Recoules \textit{et al.} \cite{Recoulesetal} and with the experimental results \cite{Brandt,Iida}.
Similarly, our results for the reflectivity agree very well with the experimental measurement near 1 eV reported in \cite{Akashev} (the experimental point is shown with the error bar representing the dispersion over the experimental frequency range).
By extending the previous Kohn-Sham calculations to higher temperature, we find a minimum in the electrical conductivities at about 6 eV, while the reflectivity coefficients show an exponential decrease with temperature. 
Similar trends are seen for the $\sigma$ and $R$ at 2.7 g/cm$^3$.

\begin{figure}[t]
  \centering
  \includegraphics[width=\columnwidth]{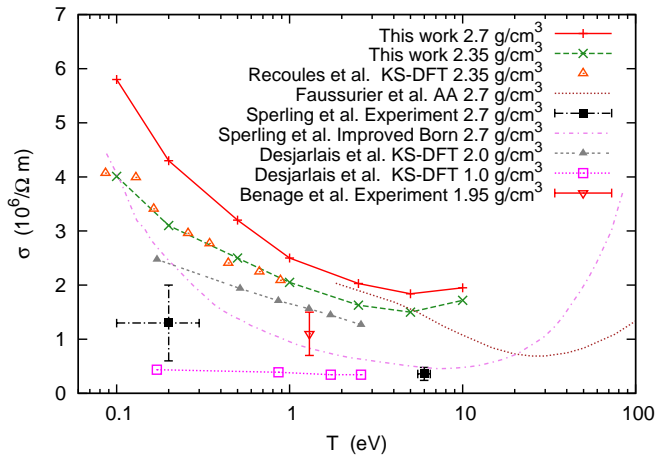}
  \caption{Electrical conductivity of aluminum. Recent experimental results for 2.7 g/cm$^3$ \cite{Sperling} are not consistent with DFT calculations and other experimental measurments.}
\label{fig:Alcond100}
\end{figure}
Surprisingly, while our results at 2.35 g/cm$^3$ are in agreement with previous experiments, our electrical conductivities at 2.7 g/cm$^3$ are inconsistent with the experimental determination recently reported by Sperling \textit{et al.} at temperatures 0.2 eV and 6 eV \cite{Sperling} (as reported in the text of their paper).
This is clearly shown in Fig. \ref{fig:Alcond100} where we reproduce the results of Fig. \ref{fig:Alcond} for the electrical conductivity together with the result of Sperling \textit{et al.} and other results discussed below.
The ``improved Born model'' calculation is also reported in Sperling \textit{et al.} and, while it was described there as in satisfying agreement with the experiment, this model is also in disagreement with our calculations.
Our calculations are also inconsistent with the recent results of Faussurier \textit{et al.} \cite{Faussurier} based on an average-atom (AA) model, which find a minimum around 20-30 eV, significantly higher to the value of $\sim$6  eV that we find.

On the other hand, our results are consistent with other Kohn-Sham calculations at 1 and 2 and 2.35  g/cm$^3$ by Desjarlais \textit{et al.} \cite{Desjarlaisetal} and Recoules \textit{et al.} \cite{Recoulesetal}, which show a clear trend of increasing conductivity with density in this temperature region.
To the contrary, the lower temperature point of Sperling \textit{et al.} even at full extent of the error bars falls between the 1 and 2 g/cm$^3$ DFT calculations.
Additionally, an experimental point from the exploding wire experiments of Benage \textit{et al.} \cite{Benage} at 1.95 g/cm$^3$, shows a conductivity a little lower than suggested by the DFT but within the error.

\section{Conclusion}

We have applied our recently-published orbital-free approximation of finite-temperature density functional theory to the calculation of ionic and electronic transport properties of dense plasmas from cold to hot conditions.
We have shown that our approach retains the level of accuracy of the orbital-based, Kohn-Sham calculations, as was previously shown for the static properties \cite{sjostromdaligault}.
Moreover, the reduction of the temperature bottleneck which exists in the Kohn-Sham method allows us to calculate these properties at a fraction of the computational cost of Kohn-Sham calculations, and further to complete calculations where Kohn-Sham is simply computationally prohibited.

The present results lend support to orbital-free quantum molecular dynamics as a viable approach that can significantly contribute to the theoretical exploration of matter under extreme conditions, especially when thermodynamic and transport properties are needed over a wide range of physical conditions of temperatures and densities.
Further work is needed to develop this approach to its full potential.
Indeed, at present, our own orbital-free method is limited to conditions where the gradients in the electron density are small enough to be consistent with the assumption made in the construction of the density functional (see discussion in Ref.\cite{sjostromdaligault}).
As a consequence, its applicability is limited to ``simple'' enough systems, such as hydrogen plasmas at large enough densities ($>$ 2 g/cm$^3$), dense aluminum below 10 eV (which can be modeled with a 3-electron pseudopotential), or very hot plasmas in which case the Thomas-Fermi approximation is adequate.
Consideration of more ``difficult'' conditions involving larger density gradients, such as aluminum calculations to temperatures above 10 eV, or of other elements that are less free-electron like, require further developments of orbital-free functionals. 
Potential research areas include the search for advanced orbital-free functionals of higher order in the density gradients \cite{Karasiev2013}; the development of accurate local pseudopotentials, as the transferability of the current pseudopotentials across densities and temperatures remains an issue, and for some elements and conditions they have been simply unattainable; the development of density functionals with a density decomposition, such as has been explored for transition metals at zero temperature \cite{HuangCarter2012}.

Further, here we have made use of the highly accurate electron density available in our orbital-free DFT to determine the Kohn-Sham potential and then obtain the Kohn-Sham orbitals by a single diagonalization which significantly reduces the computational time for high temperature calculations of the electrical and thermal conductivities through the Kubo-Greenwood linear response theory. Of course, this reliance on a calculation of Kohn-Sham orbitals for the evaluation of the conductivities is now the limiting factor. Development of an orbital-free approach which does not resort to calculating the Kohn-Sham orbitals, but retains the Kubo-Greenwood accuracy would be a significant advancement for the field, from computational and aesthetic viewpoints.

This work was carried out under the auspices of the National Nuclear Security Administration of the U.S. Department of Energy (DOE) at Los Alamos under Contract No. DE-AC52-06NA25396. The work was supported by the DOE Office of Fusion Energy Sciences.

\appendix
\section{Convergence Issues} \label{appendix1}
Here we present some of the convergence issues which were briefly stated in the main text. One such issue the necessary smearing of the $\delta$ function in Eq. \ref{eq:Ons}, which we approximate as a Lorentzian, with width $\Gamma$
\begin{align}
  \delta(\varepsilon_{\mathbf{k}\mu}-\varepsilon_{\mathbf{k}\nu}-\hbar\omega) \approx \frac{\Gamma}{\pi\left[ (\varepsilon_{\mathbf{k}\mu}-\varepsilon_{\mathbf{k}\nu}-\hbar\omega)^2 + \Gamma^2 \right]}\;.
\end{align}
\begin{figure}
  \centering
  \includegraphics[width=\columnwidth]{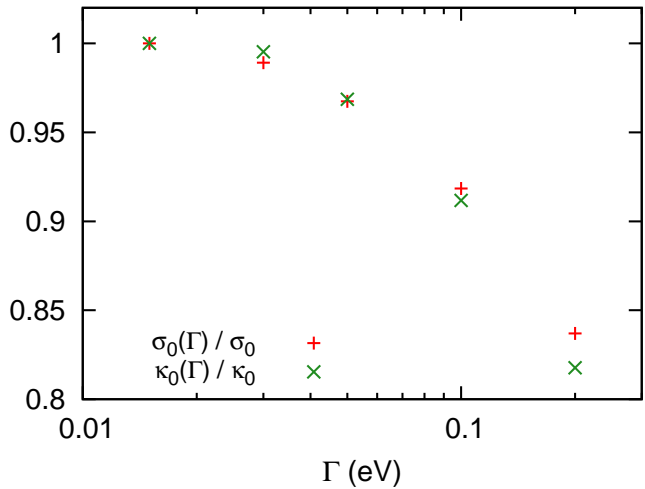}
 \caption{Convergence of the electrical and thermal conductivity for aluminum at 2.7 g/cm$^3$ and $T=5$ eV with respect to the width of the Lorentzian smearing, $\Gamma$. The value of the $\sigma_0$ and $\kappa_0$ relative to the converged values are shown.}
\label{fig:Alconv}
\end{figure}
In Fig. \ref{fig:Alconv} the convergence of the electrical and thermal conductivities are examined with respect to $\Gamma$ for the case of aluminum at 2.7 g/cm$^3$ and 5 eV.

In the case of hydrogen, we found 96 atoms sufficient down to our lowest temperatures at all densities. However, a 3 $\times$ 3 $\times$ 3 $k$-grid was needed up to 20,000 K at 10 g/cm$^3$, but only up to 5,000 and 2,000 K at 5 and 2 g/cm$^3$, and at 1 g/cm$^3$ at 1000 K a 2 $\times$ 2 $\times$ 2 $k$-grid was sufficient. We were able to use only the $\Gamma$-point for temperatures down to 232,000 K at 1 and 2 g/cm$^3$ and down to 1,000,000 K at 5 and 10 g/cm$^3$.

For the hydrogen conductivity calculations we included 450 bands at all densities and 1,000 K. While at 1,000,000 K we used 3,000 bands at 10 g/cm$^3$ and 5,000 bands at 1 g/cm$^3$.
 
\section{Calculation of the reflectivity} \label{appendix2}

Following Ref. \cite{Holst08}, with the real part of the frequency-dependent electrical conductivity  $\sigma(\omega)$ from Eq.  (\ref{eq:etrans}), the the imaginary part may be calculated through the Kramers-Kroning relation,
\begin{align}
  \sigma_2(\omega) = \frac{-2}{\pi}\;\mathrm{P} \int \frac{ \sigma_1(\nu) \omega }{\nu^2-\omega^2 } d\nu\;.
\end{align}
The real and imaginary part of the dielectric function are then given directly as
\begin{align}
  \epsilon_1(\omega) = 1 - \frac{\sigma_2(\omega)}{\epsilon_0 \omega}\;, \quad
  \epsilon_2(\omega) = \frac{\sigma_1(\omega)}{\epsilon_0 \omega}\;
\end{align}
respectively.
Then finally the real $n$ and imaginary $k$ parts of the index of refraction are found through
\begin{align}
  n(\omega) = \sqrt{\left[ |\epsilon(\omega)| + \epsilon_1(\omega) \right] / 2}\;, \\
  k(\omega) = \sqrt{\left[ |\epsilon(\omega)| - \epsilon_1(\omega) \right] / 2}\;, \\
\end{align}
and used to determine the reflectivity $R=r(\omega=0)$, where
\begin{align}
  r(\omega) = \frac{ [1 - n(\omega)]^2 + k(\omega)^2 }{ [1 + n(\omega)]^2 + k(\omega)^2 }\;.
\end{align}


\begin{thebibliography}{99}
\bibitem{notepasturel} We have also run PAW Kohn-Sham calculations using both PBE and LDA at 1000 K, and found them in agreement with the results of Ref. \onlinecite{JaksePasturel}.
\bibitem{notelambertetal}
It is of note that Ref. \cite{Lambertetal} used the Thomas-Fermi approximation to run orbital-free molecular dynamics simulations for hot dense systems, and after generating the ion configurations, they performed self-consistent PAW Kohn-Sham calculations using only the ion positions data from the Thomas-Fermi MD. Here we have gone a step further in alleviating the Kohn-Sham bottleneck, by using the Kohn-Sham potential from the orbital-free DFT and performing a non-self-consistent diagonalization.
\end{thebibliography}
\end{document}